\input harvmac
\input epsf
\noblackbox
\def\npb#1#2#3{{\it Nucl.\ Phys.} {\bf B#1} (19#2) #3}
\def\plb#1#2#3{{\it Phys.\ Lett.} {\bf B#1} (19#2) #3}
\def\prl#1#2#3{{\it Phys.\ Rev.\ Lett.} {\bf #1} (19#2) #3}
\def\prd#1#2#3{{\it Phys.\ Rev.} {\bf D#1} (19#2) #3}
\def\mpla#1#2#3{{\it Mod.\ Phys.\ Lett.} {\bf A#1} (19#2) #3}
\def\cmp#1#2#3{{\it Commun.\ Math.\ Phys.} {\bf #1} (19#2) #3}
\def\atmp#1#2#3{{\it Adv.\ Theor.\ Math.\ Phys.} {\bf #1} (19#2) #3}
\def\jhep#1#2#3{{\it JHEP\/} {\bf #1} (19#2) #3}
\newcount\figno
\figno=0
\def\fig#1#2#3{
\par\begingroup\parindent=0pt\leftskip=1cm\rightskip=1cm\parindent=0pt
\baselineskip=11pt
\global\advance\figno by 1
\midinsert
\epsfxsize=#3
\centerline{\epsfbox{#2}}
\vskip 12pt
{\bf Fig.\ \the\figno: } #1\par
\endinsert\endgroup\par
}
\def\figlabel#1{\xdef#1{\the\figno}}
\def\encadremath#1{\vbox{\hrule\hbox{\vrule\kern8pt\vbox{\kern8pt
\hbox{$\displaystyle #1$}\kern8pt}
\kern8pt\vrule}\hrule}}

\def\frac#1#2{{#1 \over #2}}

\def\semi{\subset\kern-1em\times\;}
\def\bar#1{\overline{#1}}
\def\CA{{\cal A}}

\def\CV{{\cal V}}                   
                   \def\CS{{\cal S}}

\def\R{{\bf R}}                     
                     \def\Z{{\bf Z}}
%

\Title{\vbox{\baselineskip12pt
\hbox{hep-th/0001143}
\hbox{CALT-68-2254}
\hbox{CITUSC/00-003}
\hbox{EFI-99-49}
\hbox{NSF-ITP-00-02}
\vskip-.5in}}
{\vbox{\centerline{D-Sphalerons} 
\bigskip
\centerline{and the Topology of String Configuration Space}}}
\medskip\bigskip
\centerline{Jeffrey A. Harvey$^1$, Petr Ho\v rava$^2$ and Per Kraus$^{1,3}$}
\bigskip\medskip
\centerline{\it $^1$Enrico Fermi Institute, University of Chicago, 
Chicago, IL 60637, USA}
\centerline{\tt harvey, pkraus@theory.uchicago.edu}
\medskip
\centerline{\it $^2$CIT-USC Center for Theoretical Physics}
\centerline{\it California Institute of Technology, Pasadena, CA 91125, USA}
\centerline{\tt horava@theory.caltech.edu}
\medskip
\centerline{\it $^3$Institute for Theoretical Physics, University of 
California}
\centerline{\it Santa Barbara, CA 93106, USA}
\baselineskip18pt
\medskip\bigskip\medskip\bigskip\medskip
\baselineskip16pt
We show that  unstable D-branes play the role of ``D-sphalerons'' in string 
theory.  Their existence implies that the configuration space of Type II 
string theory has a complicated homotopy structure, similar to that of an 
infinite Grassmannian.  In particular, the configuration space of Type IIA 
(IIB) string theory on $\R^{10}$ has non-trivial homotopy groups $\pi_k$ for 
all $k$ even (odd).  
\Date{January 2000}
\lref\sphal{A.~I.~Bochkarev and M.~E.~Shaposhnikov, 
``Anomalous Fermion Number Nonconservation At High Temperatures: 
Two-Dimensional Example,'' \mpla{2}{87}{991}; 
D.~Y.~Grigorev and V.~A.~Rubakov, ``Soliton Pair Creation At Finite 
Temperatures.  Numerical Study In (1+1)-Dimensions,'' \npb{299}{88}{67}.} 
\lref\sena{A.~Sen, ``Stable non-BPS bound states of BPS D-branes,''
\jhep{9808}{98}{010}, hep-th/9805019; 
``SO(32) spinors of type I and other solitons on brane-antibrane pair,''
\jhep{9809}{98}{023}, hep-th/9808141;
``Type I D-particle and its interactions,''
\jhep{9810}{98}{021}, hep-th/9809111; 
``Non-BPS states and branes in string theory,'' 
hep-th/9904207, and references therein.}
\lref\sennon{A. Sen, ``BPS D-branes on non-supersymmetric cycles,'' 
\jhep{9812}{98}{021}, hep-th/9812031.}
\lref\bergman{O.~Bergman and M.~R.~Gaberdiel,
``Stable non-BPS D-particles,'' \plb{441}{98}{133}, hep-th/9806155.}
\lref\brill{D.~Brill and G.~T.~Horowitz, ``Negative energy in string theory,'' 
\plb{262}{91}{437}.}
\lref\romans{L.~J.~Romans, ``Massive N=2a Supergravity In Ten Dimensions,'' 
\plb{169}{86}{374}.}
\lref\polchstrom{J.~Polchinski and A.~Strominger, ``New Vacua for Type II 
String Theory,'' \plb{388}{96}{736}, hep-th/9510227.}
\lref\polchinski{J.~Polchinski, ``Dirichlet-Branes and Ramond-Ramond 
Charges,'' \prl{75}{95}{4724}, hep-th/9510017.}
\lref\bergshoeff{E.~Bergshoeff, M.~de Roo, M.B.~Green, G.~Papadopoulos and 
P.K.~Townsend, ``Duality of Type II 7-branes and 8-branes,'' 
\npb{470}{96}{113}, hep-th/9601150.}
\lref\green{M.B.~Green, C.M.~Hull and P.K.~Townsend, ``D-Brane Wess-Zumino 
Actions, T-Duality and the Cosmological Constant,'' \plb{382}{96}{65}, 
hep-th/9604119.}
\lref\wittenglobal{E.~Witten, ``Global Gravitational Anomalies,'' 
\cmp{100}{85}{197}.}
\lref\alvarez{L.~Alvarez-Gaum\'e and E.~Witten, ``Gravitational Anomalies,'' 
\npb{234}{84}{269}.}
\lref\senspinors{A.~Sen, ``$SO(32)$ Spinors of Type I and Other Solitons on 
Brane-Antibrane Pair,'' \jhep{9809}{98}{023}, hep-th/9808141.}
\lref\taubes{C.H. Taubes, ``The Existence of a Nonminimal Solution to the 
$SU(2)$ Yang-Mills-Higgs Equations on $R^3$,'' \cmp{86}{82}{257}; 
{\bf 86} (1982) 299.}
\lref\manton{N.S. Manton, ``Topology in the Weinberg-Salam Theory,'' 
\prd{D28}{83}{2019}; 
F.R. Klinkhamer and N.S. Manton, ``A Saddle-Point Solution in the 
Weinberg-Salam Theory,'' \prd{30}{84}{2212}.}
\lref\baryons{V.A. Kuzmin, V.A. Rubakov and M.E. Shaposhnikov, ``On Anomalous 
Electroweak Baryon-Number Non-Conservation in the Early Universe,'' 
\plb{155}{85}{36}.}
\lref\condit{J. Cardy, ``Boundary Conditions, Fusion Rules, and the Verlinde
Formula,'' \npb{324}{89}{581};\hfill\break
P. Ho\v rava, ``Strings on World-Sheet Orbifolds,'' \npb{327}{89}{461};\hfill 
\break
``Open Strings from Three Dimensions: Chern-Simons-Witten Theory on 
Orbifolds,'' (Prague, 1990), {\it J. Geom.\ Phys.} {\bf 21} (1996) 1, 
hep-th/9404101.}
\lref\dlsfp{M. Dine, O. Lechtenfeld, B. Sakita, W. Fischler and J. Polchinski, 
``Baryon Number Violation at High Temperature in the Standard Model,'' 
\npb{342}{90}{381}.}
\lref\ewkk{E. Witten, ``Instability of the Kaluza-Klein Vacuum,'' 
\npb{195}{82}{481}.}
\lref\gpy{D.J. Gross, M.J. Perry and L.G. Yaffe, ``Instability of Flat Space 
at Finite Temperature,'' \prd{25}{82}{330}.}
\lref\theuse{S. Coleman, ``The Uses of Instantons,'' 
in {\it Aspects of Symmetry\/} (Cambridge University Press, 1985).}
\lref\ewk{E. Witten, ``D-Branes and K-Theory,'' \jhep{9812}{98}{019}; 
hep-th/9810188.}
\lref\phk{P. Ho\v rava, ``Type IIA D-Branes, K-Theory, and Matrix Theory,'' 
\atmp{2}{99}{1373}, hep-th/9812135.}
\lref\mm{R. Minasian and G. Moore, ``K-Theory and Ramond-Ramond Charge,'' 
\jhep{9711}{97}{002}, hep-th/9710230.}
\lref\yi{P. Yi, ``Membranes from Five-Branes and Fundamental Strings from
D$p$-Branes,'' \npb{550}{99}{214}; hep-th/9901159.}
\lref\senpuz{A. Sen, ``Supersymmetric World-volume Action for Non-BPS
D-branes,'' \hfill\break hep-th/9909062.}
\lref\infgenus{G. Segal and G. Wilson, ``Loop groups and equations of
KdV type,'' {\it IHES Publ.\ Math.} {\bf 61} (1985) 5; 
D. Friedan and S.H. Shenker, ``The integrable analytic geometry
of quantum string,'' \plb{175}{86}{287}; L. Alvarez-Gaum\'e, C.
Gomez and C.Reina, ``Loop groups, Grassmannians and String Theory,''
\plb{190}{87}{55}; C. Vafa, ``Operator formalism on
Riemann surfaces,'' \plb{190}{87}{47};
A. Schwarz, ``Fermionic String and Universal Moduli Space,'' 
\npb{317}{89}{323}; ``Grassmannian and String Theory,'' \cmp{199}{98}{1}, 
hep-th/9610122.}
\lref\hhktwo{J.A. Harvey, P. Ho\v rava and P. Kraus, work in progress.}
\lref\hstrom{G. T. Horowitz and A. Strominger, ``Black Strings and
$p$-Branes,'' \npb{360}{91}{197}.}
\lref\divec{P. Di Vecchia, M. Frau, I. Pesando, S. Sciuto, A. Lerda and
R. Russo, ``Classical p-Branes from Boundary State,'' \npb{507}{97}{259}.}
\lref\moorwit{G. Moore and E. Witten, ``Self-Duality, Ramond-Ramond Fields,
and K-Theory,'' hep-th/9912279.}
\lref\yoneya{T.Yoneya, ``Spontaneously Broken Space-Time Supersymmetry in
Open String Theory Without GSO Projection,'' hep-th/9912255.}
\lref\senzw{A. Sen and B. Zwiebach, ``Tachyon Condensation in String Field 
Theory,'' hep-th/9912249.}
\lref\bcr{M. Bill\'o, B. Craps and F.Roose, ``Ramond-Ramond couplings of
non-BPS D-branes,'' \jhep{9906}{99}{033}; hep-th/9905157.}
\newsec{Introduction}

Most of the recent progress in non-perturbative string theory has been 
facilitated by the powerful constraints imposed by supersymmetry.  It seems  
vital, for both theoretical and phenomenological reasons, to extend our 
understanding to configurations where some of these constraints have been 
relaxed.  Non-supersymmetric string vacua are of course notoriously difficult 
to study.  It seems reasonable, therefore, to first analyze non-supersymmetric 
excitations in the supersymmetric vacua of the theory.  

During the last year it has been realized that the spectrum in some vacua of 
string theory contains not only BPS D-branes, but also stable non-BPS D-branes 
\refs{\sena,\bergman,\ewk}.  
A very useful perspective for the study of these non-supersymmetric 
D-brane configurations has been developed by Sen \sena .  In this framework, 
one views stable D-branes as bound states on the worldvolume of an unstable 
system composed of BPS D-branes and anti-D-branes with a higher worldvolume 
dimension.  This construction has been further generalized \refs{\ewk,\phk}, 
leading to a systematic framework which implies that D-brane charges on a 
compactification manifold $X$ are classified by a generalized cohomology 
theory of $X$ known as K-theory as suggested by previous work on
Ramond-Ramond charges \mm .  

A crucial role in this framework is played by unstable D$p$-brane systems.  
In dimensions where RR-charged D$p$-branes exist, one can construct unstable 
systems by considering D$p$-D$\bar p$ pairs.  For the ``wrong'' values of $p$, 
where stable RR-charged D$p$-branes do not exist, it was realized 
\refs{\sennon,\phk} that one can still construct an {\it unstable\/} 
D$p$-brane.%
\foot{Such unstable D-branes (in particular, the spacetime-filling D9-branes) 
are indeed crucial in the systematic classification of D-brane charges in Type 
IIA theory and its relation to $K^{-1}(X)$ groups of spacetime \phk .}
Thus, in addition to the RR-charged BPS D-branes, there are unstable 
D$p$-branes for $p$ odd in Type IIA theory, and $p$ even in Type IIB theory.  

Such unstable D-branes can be directly constructed in the boundary-state 
formalism.  Consider Type IIA or IIB theory 
in $\R^{10}$.%
\foot{In this paper, we focus on Type IIA and Type IIB string theory in 
$\R^{10}$, making only occasional comments about orientifolds and 
compactifications.}
The boundary state describing a D$p$-brane can have a contribution from the 
closed string 
NS-NS sector and  RR sector.  {}For each $p$, there is a unique boundary 
state in the NS-NS sector that implements the correct boundary conditions, 
and survives the corresponding GSO projection.  The RR sector, on the other 
hand, contains a unique boundary state only for those D$p$-branes that 
can couple to a RR form $C_{p+1}$; for all other values of $p$, the GSO 
projection kills all possible boundary states in the RR sector.  

The supersymmetric RR-charged D$p$-brane is  described by 
\eqn\eebcsusy{\left| {\rm D}p\right\rangle_{\rm BPS}=\frac{1}{\sqrt{2}}\left(
\left| B\right\rangle_{\rm NS\,NS}\pm\left| B\right\rangle_{\rm R\,R}\right),}
where the sign in front of the RR component of the boundary state determines 
the RR charge of the brane.  

In contrast, the boundary state describing the D$p$-brane for the ``wrong'' 
values of $p$ -- where $C_{p+1}$ is absent --  contains only a  NS-NS 
component, 
\eqn\eebc{\left| {\rm D}p\right\rangle= \left| B\right
\rangle_{\rm NS\,NS}.}
The relative factor of $\sqrt{2}$ and the consistency of this set of
boundary states follows from constructing the cylinder amplitudes with
all possible pairs of boundary states and imposing the condition that the 
cylinder amplitude has a consistent open string interpretation 
\refs{\condit}. 

The spectrum of open strings ending on unstable D-branes is non-supersymmetric 
and contains a tachyon.  To see this, note that the absence of the RR sector 
in the boundary state implies the absence of the GSO projection in the 
open-string loop channel, and as a result, the 
open-string spectrum  contains {\it both\/} the lowest tachyonic mode $T$ 
{\it and\/} the gauge field $A_M$.  {}For $N$ coincident unstable D-branes in 
Type II theory, the gauge symmetry is $U(N)$, and the tachyon $T$ is in the 
adjoint representation of $U(N)$. 

The unstable D$p$-branes with worldvolumes of the ``wrong'' dimension 
represent legitimate classical solutions of open string theory, despite the 
fact that they are non-supersymmetric, unstable, and carry no charge.  And, 
as for BPS D-branes, one expects these solutions to be a good approximation 
to solutions of the full closed and open string theory for small string 
coupling.  The present paper is devoted to clarifying the physical 
interpretation of 
the unstable D-branes in string theory.  

To whet the reader's appetite we offer the following observation. In Type IIA 
theory, unstable D$p$-branes exist for $p$ odd and, in particular, there is a 
Type IIA  D-instanton.  This D-instanton represents a Euclidean solution of 
the theory with a fluctuation spectrum containing one negative eigenvalue.  
Instantons with exactly one negative eigenvalue often represent a ``bounce,'' 
or false vacuum decay; the square root of the fluctuation
determinant is imaginary due to the single negative eigenvalue and the
imaginary part of the vacuum amplitude gives the vacuum decay rate. 
For a review see \theuse. 
In higher-dimensional theories with gravity, such vacuum decay often has 
disastrous consequences, 
leading to a complete annihilation of spacetime that starts by nucleation of 
a hole which then expands with a speed approaching the speed of light \ewkk.  

These observations lead to an intriguing question: does the existence of a 
D-instanton with one negative eigenvalue in Type IIA theory signal that 
its supersymmetric vacuum is false, and therefore unstable to decay?  Before 
jumping to conclusions, declaring that the supersymmetric Type IIA vacuum 
(and, by duality, perhaps all other supersymmetric vacua) is unstable, and 
interpreting this as a string phenomenologist's dream, one needs to carefully 
examine whether the Type IIA D-instanton represents a bounce for false vacuum 
decay.  

In an attempt to answer this question, we will clarify the role of all the 
unstable D$p$-branes.  In particular, we will see that the Type IIA 
D-instanton does not represent a bounce signaling an instability of the 
supersymmetric Type IIA vacuum.  Instead, the D-instanton is tied to a 
completely different physical phenomenon, also with a precedent in field 
theory.  We will find 
that the unstable D-branes in superstring theory are intimately related to the 
surprisingly complicated topological structure of the configuration  space 
of string theory.  In field theory, classical solutions 
with a negative mode that are mandated by non-trivial homotopy of the 
configuration space are called sphalerons.  The main observation of this 
paper is that the unstable D-branes are precise string-theoretical analogs 
of sphalerons of field theory; we hope to convince the reader that it makes 
sense to call them D-sphalerons.  Thus, the spectrum of D-branes in Type IIA 
theory consists of D$(2p+1)$-brane sphalerons (``D$(2p+1)$-sphalerons'' for 
short) and BPS D$2p$-branes, while the Type IIB spectrum contains 
D$2p$-sphalerons and BPS D$(2p+1)$-branes. We will see that the existence of 
the D-sphalerons follows from the fact that the configuration space of IIA 
(IIB) string theory in $\R^{10}$ has nontrivial homotopy groups $\pi_{k}$ for 
all $k$ even (odd), and is thus homotopically at least as complicated as 
an infinite Grassmannian (the infinite unitary group).  
 
\newsec{Unstable D-branes}

In our discussion, it will be convenient to  use interchangeably several
different representations of D$p$-branes, which we first review.  

\item{(i)} The traditional representation, as a hypersurface $\Sigma_{p+1}$ 
in spacetime where fundamental strings can end.  In string perturbation 
theory, D-brane dynamics is described by open strings ending on the brane; the 
boundary conditions are summarized by the closed-string boundary states 
\eebcsusy\ and \eebc .  The unstable D-brane carries no charge, and the only 
long-distance fields associated with it are the dilaton and
graviton;  the theory is in its 
supersymmetric vacuum in the regions far away from $\Sigma_{p+1}$.   

\item{(ii)} The topological defect representation, as a bound state extended 
along a submanifold $\Sigma_{p+1}$ inside the worldvolume $\Sigma_{q+1}$ of an 
unstable D$q$-brane system with $q>p$.  

\item{(iii)} The spacetime representation in terms of a solution to the
closed string equations of motion. This is well understood for BPS
D-branes \refs{\hstrom, \divec} and will be partially developed in what
follows for non-BPS D-branes.   

\noindent
There are two unstable D-brane systems relevant  for the construction in 
point~(ii).  
When $q$ is such that RR-charged D$q$-branes exist, the unstable system is 
given by $N$ D$q$-D$\bar q$ pairs.  For the complementary values of $q$, the 
unstable system is simply the set of $2N$ unstable D$q$-branes of \eebc .  

In both cases, the worldvolume theory on $\Sigma_{q+1}$ contains a tachyon 
field $T$ which behaves as a Higgs field, rolling down to the minimum of its 
potential and Higgsing the gauge symmetry on $\Sigma_{q+1}$.  The structure 
of the gauge symmetries and the symmetry breaking patterns are summarized in 
the following table: 
\eqn\eetabone{\vbox{\offinterlineskip \hrule
\halign{&\vrule#&\strut\ \ \hfil#\ \ \cr
height2pt&\omit&&\omit&&\omit&&\omit&\cr 
&unstable system:\ &&gauge symmetry:&&tachyon:&&vacuum manifold:\quad&\cr
\noalign{\hrule}
height2pt&\omit&&\omit&&\omit&&\omit&\cr
&$N$ D$q$-D$\bar q$ pairs\ \ &&$U(N)\times U(N)$&&$(N,\bar N)\ $&&$U(N)\qquad
\quad$&\cr 
\noalign{\hrule}
height2pt&\omit&&\omit&&\omit&&\omit&\cr 
&$2N$ unstable D$q$'s&&$U(2N)\qquad$&&adjoint
&&$U(2N)/U(N)\times U(N)$&\cr}\hrule}}
Notice that in the case of the unstable D$q$-branes, the correct spectrum 
of stable D$p$-branes as defects in flat spacetime is reproduced by the 
symmetric Higgs pattern, with $U(2N)$ broken to $U(N)\times U(N)$ 
\phk .  The role of configurations with an odd number of unstable D$q$-branes, 
as well as asymmetric Higgs patterns, will be discussed in section~6.  

The Higgs mechanism, whereby the tachyon uniformly condenses to the minimum of 
its potential, can be thought of as the worldvolume representation of how the 
unstable brane system decays to the vacuum.  This interpretation of the 
Higgs mechanism leaves one obvious puzzle: the existence of the residual 
gauge symmetry, which should be absent in the true supersymmetric vacuum of 
the theory.  Various attempts to resolve this puzzle have been proposed in the 
literature \refs{\yi,\senpuz}.  In this paper, we will not address this issue, 
and will simply {\it assume\/} that the unstable D-brane system with 
the tachyon uniformly condensed to the minimum of its potential is nothing but 
a somewhat awkward representation of the supersymmetric vacuum of the theory.%
\foot{Strong evidence supporting this assumption has been recently obtained, 
with the use of string field theory, by Sen and Zwiebach \senzw\ in the 
closely related case of an unstable D-brane in the bosonic string.}

The unstable D-brane systems \eetabone\ support a host of topological 
defects, which are interpreted as lower-dimensional D-branes.  Stable defects 
are classified by non-trivial elements in the homotopy groups of the vacuum 
manifolds, 
\eqn\eehomvone{\eqalign{\pi_{2k+1}(U(N))&=\Z,\cr\pi_{2k}(U(N))&=0,\cr}}
and 
\eqn\eehomvtwo{\eqalign{\pi_{2k+1}(U(2N)/U(N)\times U(N))&=0,\cr
\pi_{2k}(U(2N)/U(N)\times U(N))&=\Z.\cr}}
(These formulas hold in the stable regime, of $N$ sufficiently large for 
fixed $k$.) These homotopy groups are directly related to K-theory groups of 
spacetime \refs{\phk,\ewk}; therefore, D-brane charges are naturally described 
in K-theory.  

First, consider a BPS D$p$-brane.  This brane can be represented as a 
codimension $p'$ defect along $x^i=0,i=1,\ldots p'$, in an unstable system of 
D$q$-branes with $q=p+p'$.  For a codimension $p'$ defect, the tachyon field 
$T$ maps the sphere $S^{p'-1}$ at infinity in the transverse dimensions to the 
vacuum manifold $\CV$, thus defining an element of $\pi_{p'-1}(\CV)$.  
In even codimension $p'=2k$, the unstable system consists of $N=2^{k-1}$ 
D$q$-D$\bar q$ pairs, and in odd codimension $p'=2k-1$, of $N=2^{k-1}$ 
unstable D$q$-branes. The corresponding tachyon condensate is given 
explicitly by 
\eqn\eegamma{T=f(r)\Gamma_i x^i,}
where $\Gamma_i$ are the gamma matrices of the rotation group $SO(p')$ in the 
transverse dimensions $x^i$.%
\foot{The convergence factor $f(r)$ only depends on the radial coordinate, and 
asymptotes to $T_0/r$ as $r\rightarrow\infty$, with $T_0$ one of the 
eigenvalues of $T$ at the minimum of its potential; $f(0)=1$.  This 
convergence factor will be systematically omitted throughout the paper.}
The gauge field $A_M$ on the D$q$-brane system is also non-zero, such that 
the energy of the whole configuration is finite. 

For even $p'$, we will have occasion to use two distinct definitions of 
the gamma matrices.  Let $\CS_+$ and $\CS_-$ be the two $2^{n-1}$ dimensional 
irreducible spinor representations of $SO(2n)$.  We can either define 
$\Gamma_i$ to be $2^{n-1} \times 2^{n-1}$ matrices mapping $\CS_+$ to 
$\CS_-$, or to be $2^{n} \times 2^{n}$ matrices mapping $\CS_+ \oplus \CS_-$ 
to itself.   Which definition is being used will be always be 
implied by the stated dimensionality of the matrices.

Now, consider an unstable D$p$-brane, described by the boundary state \eebc .  
The tachyon field
now defines a homotopically trivial map to the vacuum manifold which is
reflected by the instability of the D$p$-brane. However we still expect a 
solution with the core carrying a finite energy density along the worldvolume
$\Sigma_{p+1}$. We will argue that this brane is also described by the same 
formula \eegamma , as a defect of codimension $p'$  in a corresponding 
unstable brane system, even though there is no direct topological argument as 
there is for BPS D-branes.  

Consider as an  example a D$p$-brane with $p$ even. In IIA theory this is a 
stable BPS brane and may be represented in various ways as a topological defect
in higher dimensional  unstable brane systems. The simplest is as a kink in
the real tachyon field of the unstable D$(p+1)$-brane of IIA.  Now we compare 
this to the unstable D$p$-brane of IIB,  taking as our starting point 
the unstable D$(p+1)$-D$\bar{(p+1)}$ system with a complex tachyon and a 
``mexican hat'' potential. 
If we can establish that a cross-section of this potential gives the 
double-well 
potential of the IIA D$p$ system, then it is clear that the previous kink 
solution is again a solution, but now with an instability due to the 
possibility of pulling the kink off the top of the potential. 

That the potential has this property follows from the rules developed in
\sena.   Open strings ending on an unstable D$p$-brane are assigned 
$2 \times 2$ Chan-Paton matrices.  The $U(1)$ gauge field  is assigned
to $1$ and the real tachyon is assigned to $\sigma_1$.   Open strings of 
the D$p$-D$\bar{p}$ system also have $2 \times 2$ Chan-Paton matrices: 
the $U(1)\times U(1)$ gauge fields are assigned to $1$ and $\sigma_3$, and
the complex tachyon is assigned to $\sigma_1$ and $\sigma_2$.   Then at
the level of disk diagrams, the action for the tachyon of the unstable
D$p$-brane is the same as for the  $\sigma_1$ component of the 
D$p$-D$\bar{p}$ tachyon.   

This line of argument actually establishes that any solution on the 
unstable D$p$-brane yields a solution on the D$p$-D$\bar{p}$ system, once the
real tachyon is mapped to the $\sigma_1$ component of the complex tachyon, and
the $U(1)$ gauge field is mapped to the $1$ component of the $U(1)\times U(1)$
gauge field.  The remaining fields associated to $\sigma_2$,
$\sigma_3$ will only appear at least quadratically in fluctuations, since
the trace over Chan-Paton matrices eliminates any linear terms.  The quadratic
fluctuations may destabilize a solution constructed in this fashion, but 
won't change the fact that it is a solution.  One can also translate 
solutions in the opposite direction; a similar argument establishes that
solutions on two coincident unstable D$p$-branes can be mapped to solutions
on a D$p$-D$\bar{p}$ system.  Again, the additional fields on the    
D$p$-D$\bar{p}$ system appear at least quadratically in fluctuations,
and so at worst destabilize the solution.    We thus conclude that the 
unstable D$p$-brane is also described by \eegamma\ as a defect on the 
worldvolume of an unstable D$q$-brane system, despite the fact that this 
configuration is topologically unstable.  

In the following we will make frequent use of the ability to translate
stable and unstable solutions in the above manner, although we
will not know the explicit solutions beyond their asymptotic behavior.      

\subsec{Type IIA D-instanton}

Equipped with these different representations of unstable D-branes, let us 
return to the Type IIA D-instanton.  For a classical Euclidean solution with 
one negative eigenvalue to represent a bounce, it has to satisfy several 
conditions.  
\fig{False vacuum decay and the Euclidean instanton (the ``bounce'') with 
one negative eigenvalue that dominates the path 
integral.}{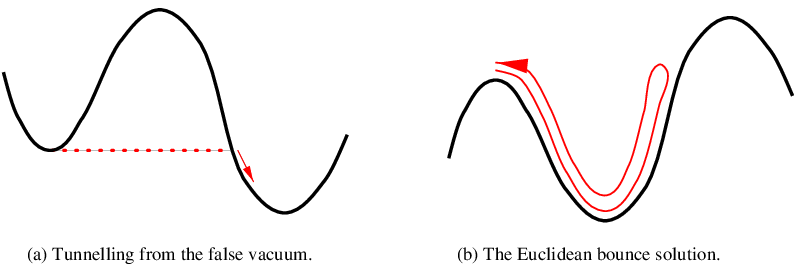}{4.5truein}
\noindent
First of all, it has to be asymptotic to 
the false vacuum in all directions.  The fate of the false vacuum after the 
tunneling can be read off from the bounce, by identifying its turning point, 
and evolving the configuration classically in the Minkowski signature.  For a 
solution to admit such a procedure, it has to have a reflection symmetry 
along a codimension-one surface, which we can identify with the surface of 
constant Euclidean time $\tau_E=0$.  Moreover, we must be able to Wick-rotate 
the solution to the Minkowski regime.  The turning point is
defined to be a point on the trajectory where the kinetic energy vanishes,
so by (Euclidean) energy conservation the potential energy at the turning
point equals the potential energy of the false vacuum.   Other points on 
the bounce trajectory have higher potential energy, they are under the barrier.
If $\Phi(\vec x,\tau_E)$
represents the bounce, and $\Phi(\vec x, -\tau_E)= \Phi(\vec x, \tau_E)$ then
the Euclidean kinetic energy vanishes at the turning point $\tau_E=0$.

Consider the Type IIA D-instanton, first in the supergravity approximation.  
The supergravity solution that represents our D-instanton should respect the 
$SO(10)$ rotation symmetry, and be asymptotic to the supersymmetric vacuum 
of Type IIA theory.  The only fields that can be excited are the metric and 
the dilaton; unlike in Type IIB theory, there is no ``axion'' that could be 
excited.  It is useful to interpret Type IIA theory as M-theory on $S^1$.  
The Type IIA dilaton is related to the 11-11 component of the 
eleven-dimensional metric.  Therefore, the only field excited in the 
D-instanton background is the eleven-dimensional metric.  The  equations of 
motion for this metric are just the vacuum Einstein equations, constrained by 
the requirement of $SO(10)$ rotation symmetry and $U(1)$ translation symmetry 
along the eleventh dimension.  By the eleven-dimensional analog of the 
Birkhoff theorem, the solution -- at least away from the D-instanton core -- 
has to be given by the Euclidean Schwarzschild metric, which in appropriate 
coordinates takes the form
\eqn\eemeteleven{ds^2=\left(1-\frac{M}{r^8}\right)(dx^{11})^2+
\frac{dr^2}{1-M/r^8}+r^2d\Omega_9^2.}
Furthermore, it is clear that this solution of M-theory represents a D-brane
in the sense of a possible end point for strings. To see
this, note that a membrane wrapped on the ``cigar'' of Euclidean Schwarzschild
represents a fundamental string far from the core, and this string 
clearly ends at the core of the solution.  

In spite of all this circumstantial evidence, the  usual smooth Euclidean
Schwarzschild solution does not correctly represent the IIA D-instanton. 
In the  solution \eemeteleven\ there are two parameters:  the ``mass'' 
parameter $M$, and 
the value of the radius of the eleventh dimension.  This is as it should be, 
because we expect two parameters in the IIA D-instanton system in the 
supergravity approximation:  the string coupling constant at infinity, and the 
number $N$ of D-instantons.  Supergravity cannot distinguish the 
discreteness of the second quantum number $N$, and sees it as a smooth 
parameter $M$.  The string coupling is related in the usual way to the radius 
of the eleventh dimension $x^{11}$, and can be adjusted arbitrarily.  This of 
course leaves a conical singularity at $r=r_0\equiv M^{1/8}$, corresponding 
to the location of the D-instanton(s) at $r=r_0$.  

As we will now explain, such a singularity is inevitably present in the 
supergravity solution describing the D-instanton.  
If we impose the additional requirement of smoothness at $r=r_0$ on 
\eemeteleven , we obtain the Euclidean Schwarzschild black hole, with the 
radius $R_{11}$ of $S^1$ uniquely determined by the parameter $M$, 
$R_{11}=M^{1/8}/4$.  So far we 
have ignored the presence of fermions in the theory.  The D-instanton is a 
non-supersymmetric solution in a supersymmetric theory, asymptotic at infinity 
to the supersymmetric vacuum.  Therefore, the spin structure it carries has 
to preserve supersymmetry asymptotically at infinity.  In the 
eleven-dimensional representation, the spin structure on the eleven-manifold 
\eemeteleven\ describing the D-instanton has to correspond to periodic 
boundary conditions on the fermions around the $S^1$.  In contrast, the 
smoothness of the Euclidean black hole implies that it can carry only one 
spin structure, with fermions antiperiodic around the $S^1$.  This in turn 
implies that the metric of the D-instanton always has to have a singularity 
at $r=r_0$, in order to carry the correct, that is periodic, spin structure.  

As we have just seen, the singularity of the metric at the location of the 
D-instanton cannot be resolved by supergravity; in particular, one cannot 
count the number of negative modes of the solution in the supergravity 
approximation.  
\fig{Two supergravity solutions: (a) The Euclidean Schwarzschild black hole 
in eleven dimensions; (b) the Type IIA D-instanton.}{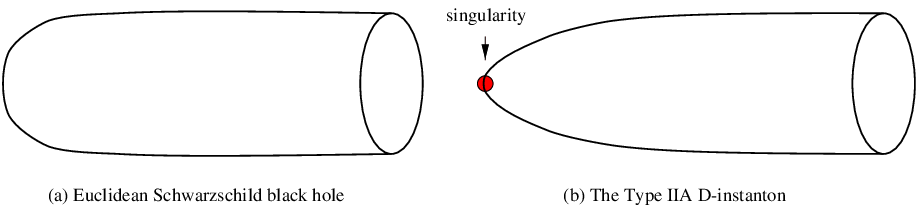}{5.5truein}
\noindent
On the other hand, the Euclidean Schwarzschild is a smooth solution, with 
only one free parameter -- the value of the string coupling at infinity, and 
will have exactly one negative mode.  Since its spin structure is that of 
antiperiodic fermions around $S^1$ at infinity, the Euclidean Schwarzschild 
represents a bounce relevant for the fate of the vacuum in a large class of 
compactifications, related to M-theory (or string theory) on $S^1$ with the 
non-supersymmetric spin structure \refs{\brill,\hhktwo}, and also describes 
black hole nucleation in M-theory at finite temperature, as in \gpy . 

The singularity found at the tip of the supergravity solution is resolved in 
string theory by the presence of the D-branes.  In this representation, one 
can count the number of negative modes of this configuration.  $N$ coincident 
D-instantons will have $N^2$ negative modes, from the open-string tachyon in 
the adjoint of $U(N)$.  Clearly, it is only the single-instanton configuration 
that can in principle represent a bounce.  Notice that $N=1$ will correspond 
to a small value of $M$, and therefore the supergravity approximation will be 
invalid for this system.

We will now analyze the turning point configuration for the
Type IIA D-instanton by 
using  the representation of the IIA D-instanton as a topological defect 
of the familiar form 
\eqn\eeinstga{T=\Gamma\cdot x,}
on 32 unstable D9-branes.  Here $\Gamma_i$ are $32 \times 32$ SO(10) gamma
matrices.  

 Consider a $9+1$ split of coordinates, 
$x^i=(\vec x, x^{10})$.  Using two equivalent representations of the $\Gamma$ 
matrices of $SO(10)$, we can write \eeinstga\ in two forms leading to two 
different physical interpretations of the D-instanton \eeinstga .  First, 
\eeinstga\ can be written as 
\eqn\eedec{T=\pmatrix{x^{10}\,1_{16}&\vec\Gamma\cdot\vec x\cr & \cr
\vec\Gamma \cdot\vec x&-x^{10}\,1_{16}\cr},}
where $\vec\Gamma$ are the gamma matrices of $SO(9)$.  This corresponds to 
first forming sixteen D8--branes and sixteen D$\bar 8$-branes as kinks 
localized at $x^{10}=0$ on 32 D9-branes, represented in \eedec\ by the terms 
along the diagonal.  The D-instanton then appears as the bound state 
$\vec\Gamma\cdot\vec x$ of sixteen D8-D$\bar 8$ pairs.  

Alternatively, one can write \eeinstga\ as 
\eqn\eedectwo{T=\pmatrix{\vec\Gamma\cdot\vec x&x^{10}\,1_{16}\cr & \cr
x^{10}\,1_{16}&-\vec\Gamma\cdot\vec x\cr}.}
In this picture, we use the D9-branes to first prepare a D0-D$\bar 0$ pair 
(represented by the diagonal terms in \eedectwo), with their worldline along 
$x^{10}$, and then form a kink along $x^{10}$ on the worldline of the 
D0-D$\bar 0$ system.  It is this latter representation \eedectwo\ of 
\eeinstga\ that is useful for determining the physical meaning of the 
``halfway point'' of the D-instanton.  Setting $x^{10}=0$ in \eedectwo\ leaves 
the configuration consisting of a D0-D$\bar 0$ pair at Euclidean time 
$x^{10}=0$.  

The D-instanton does indeed possess a reflection symmetry in $x_{10}$, 
which in the form \eedectwo\ is given by $T(\vec x, -x^{10}) = \sigma_3
T(\vec x, x^{10}) \sigma_3^{-1}$. However, because of the gauge transformation
which accompanies the reflection, this symmetry does not imply vanishing
of the kinetic energy%
\foot{Strictly speaking, we should study the vanishing
of the gauge covariant kinetic energy, but turning on a non-zero gauge field
to cancel the time derivative of $T$ will simply generate a non-zero electric
field leading to non-zero gauge kinetic energy.}
at the symmetry point $x^{10}=0$, which therefore is not a proper turning 
point.   Alternatively, we can use the decomposition \eedectwo\ of the 
D-instanton to see that an energy condition is being violated at the halfway 
point, and the instanton therefore cannot represent a legitimate bounce.  We 
have seen that the halfway point consists of a D0-D$\bar 0$ pair on top of the 
supersymmetric vacuum; however, such a configuration carries positive energy 
with respect to the supersymmetric vacuum, and its nucleation is forbidden by 
energy conservation.  We conclude that the Type IIA D-instanton does not lead 
to false vacuum decay of the supersymmetric Type IIA vacuum.  

\newsec{Unstable D-Branes as D-Sphalerons}

We have argued that the Type IIA D-instanton does not represent a bounce 
for false vacuum decay of the supersymmetric vacuum in Type IIA theory.  
In this section, we start collecting evidence leading to a different physical 
interpretation of all the unstable D-branes, and the Type IIA D-instanton in 
particular.  

\subsec{Sphalerons in field theory}

In field theory, sphalerons \manton\ are static solutions of the classical 
equations of motion with a single negative mode, 
whose existence is implied by a 
non-contractible loop in the configuration space of the theory.  
\fig{The topological argument tying the existence of a non-contractible loop 
in the configuration space with the existence of a static solution with one 
negative eigenvalue (the sphaleron).  The vertical axis corresponds to the 
energy.}{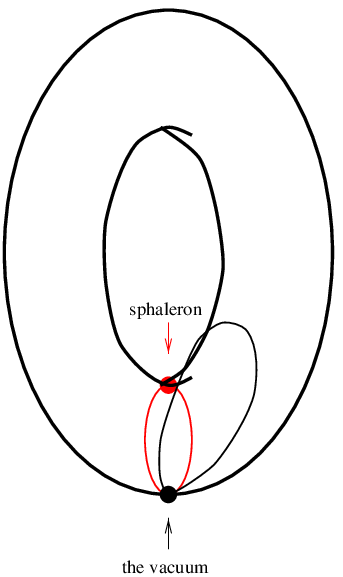}{1.5truein}
The  argument \refs{\taubes,\manton} goes as follows.  Consider 
Yang-Mills gauge theory with matter in $D+1$ spacetime dimensions.  This 
theory has a configuration space $\CA$, of all physically inequivalent, finite
energy  
configurations on the $D$-dimensional space.  Assume now that $\CA$ contains 
a non-contractible loop, i.e., that $\pi_1(\CA)\neq 0$.  If $\CA$ is 
sufficiently compact, the situation can be visualized as in Figure~3.  
Choose an arbitrary non-contractible loop $\ell$ in $\CA$ which begins and 
ends in the vacuum, and parameterize this loop by $t\in[0,2\pi]$.  Without 
any loss of generality, assume that the energy along $\ell$ grows 
monotonically as we move away from the vacuum, and reaches its absolute 
maximum $E(\ell)$ at the half-point $t=\pi$.  Since $\ell$ is 
non-contractible, there is a loop $\ell_0$ homotopically equivalent to $\ell$ 
and such that 
\eqn\eeminimum{E_0\equiv E(\ell_0)\leq E(\ell')}
for all loops $\ell'$ that are homotopically equivalent to $\ell$.  The 
point in the configuration space $\CA$ that corresponds to $t=\pi$ along 
such a minimal loop $\ell_0$ is guaranteed to be a static, finite-energy 
solution of the theory, called {\it the sphaleron}.  The spectrum of 
fluctuations around the sphaleron will contain precisely one negative 
eigenvalue, corresponding to the two directions in which the sphaleron can 
slide down to the true vacuum along the loop $\ell_0$ in the configuration 
space.  

A loop $\ell$ in the configuration space $\CA$ represents a one-parameter set 
of $D$-dimensional configurations, and can be viewed as a $D+1$-dimensional 
Euclidean configuration with the Euclidean time given by the loop parameter 
$t$.  The loop $\ell$ will be non-contractible if this $D+1$ dimensional 
configuration is topologically stable.  At infinity in all Euclidean 
dimensions, this configuration is mapped to the vacuum manifold $\CV$ of the 
theory.  Thus, the non-contractible loop determines a non-trivial element 
of $\pi_D(\CV)$.  

Notice that the sphaleron in a $D$-dimensional space carries no conserved 
topological quantum numbers, since it can be continuously connected to the 
vacuum.  In other words, the sphaleron can be unwrapped at $S^{D-1}$ at 
infinity, and 
corresponds to the trivial element in $\pi_{D-1}(\CV)$.  However, it is the 
non-contractible loop in the configuration space that is supported by the 
non-trivial element in $\pi_D(\CV)$.  In this sense, there is a certain 
similarity between instantons and the Euclidean configuration representing 
the non-contractible loop, as it is the same quantum number that is 
responsible for both.  However, even though the topology is similar, the 
energetics is different.  In the case of an instanton, we impose a single 
condition of finite action in $D+1$ dimensions, while in the case of a loop 
in configuration space, we impose the finite-energy condition in $D$ 
dimensions for each value of the loop parameter $t$.  

We now illustrate this general construction with a few simple examples:

\item{(1)} The original sphaleron of \manton\ was found in a simplified 
version of the standard model, given by the $SU(2)$ theory with a doublet 
Higgs in $3+1$ dimensions.  The vacuum manifold is a three-sphere $S^3$.  
A non-contractible loop exists, and corresponds to a point-like topological 
defect in four Euclidean dimensions that non-trivially wraps around the 
$S^3$, and corresponds to the generator in $\pi_3(S^3)=\Z$.  The sphaleron 
is an unstable static solution in the vacuum sector.  

\item{(2)} As an even simpler example consider the Abelian Higgs model,
\eqn\abhigss{
S = \int \! d^2x\, \left\{ -{1 \over 4} F_{\mu\nu}^2 + 
|(\partial_\mu - i A_\mu)\phi|^2 - {1 \over 4} \lambda(|\phi|^2-1)^2 \right\}.}
 The configuration space of this theory also has 
a non-contractible loop, given by a point-like vortex in two Euclidean 
dimensions, stable because the Higgs field at infinity corresponds to the 
generator of $\pi_1(S^1)$.  The corresponding sphaleron~\sphal\ is given by
\eqn\sphalsol{
\phi = \tanh{\left[{1 \over 2} \sqrt{\lambda}(x-x_0)\right]} e^{i\beta(x)},
\quad
A_0 = 0,
\quad
A_1 = \partial_x \beta(x),}
where $x_0$ is arbitrary and the only condition on $\beta$ is 
\eqn\betacon{
\beta(\infty) - \beta(-\infty) = \pi.}

\subsec{Unstable D0-brane as a D-sphaleron in Type IIB theory}

Consider Type IIB string theory on $\R^{10}$.  This theory has an unstable 
D-particle with a single real tachyon.  The system of $N$ such coincident 
D-particles has a tachyon in the adjoint of $U(N)$ on the worldline.  
Upon orientifold projection to Type I, the unstable Type IIB D-particle 
becomes the $\Z_2$-charged stable non-BPS D0-brane of Type I string theory 
\refs{\sena,\bergman,\ewk}. This is 
because only the antisymmetric part of the 
adjoint tachyon survives the $\Omega$ projection, leaving an
instability for $N>1$ but making the $N=1$ system 
stable.  Here, however, we are interested in the unstable D0-brane of Type 
IIB theory in its own right.  

The D0-brane of Type IIB theory can be viewed as a defect, represented by 
$\Gamma\cdot x$ of \eegamma, on sixteen D9-D$\bar 9$ pairs, where $\Gamma_i$  
are the $SO(9)$ gamma matrices of the rotation group in the nine transverse 
dimensions $x^i$.  This configuration is topologically unstable: the tachyon 
maps the 8-sphere at infinity to the vacuum manifold, but the relevant 
homotopy group $\pi_8(U(16))=0$ is trivial.  The $SO(9)$ group has only one 
spinor representation $\CS$, and the gamma matrices represent a map 
$\CS\rightarrow\CS$.  Since the $\Gamma\cdot x$ configuration carries no 
D-brane charge, it corresponds to the trivial element in K-theory.  Thus, the 
Chan-Paton bundle supported by the D9-branes is isomorphic to the Chan-Paton 
bundle of the D$\bar 9$-branes, and both are identified with $\CS$ (extended 
to the whole spacetime manifold $\R^{10}$).  

We now claim that the D0-brane is a D-sphaleron, i.e., it is a static solution 
of the equations of motion of Type IIB string theory that has one negative 
mode, and represents the top of the potential barrier along a non-contractible 
loop in the configuration space of Type IIB string theory on the non-compact 
space $\R^9$.  We will prove this directly by constructing the 
corresponding non-contractible loop in the configuration space, i.e., a 
one-parameter set of configurations on $\R^9$ (parametrized by $ t'
\in[0,2\pi]$) which begins and ends in the supersymmetric vacuum, and at 
$ t'=\pi$ passes through the configuration describing the unstable D0-brane.  

In our construction, we use the defect representation of the D0-brane, as 
$\Gamma\cdot x$ on sixteen D9-D$\bar 9$ pairs.  A one-parameter family of 
configurations on $\R^9$  can be viewed as a Euclidean configuration on 
$\R^9\times \R$, parametrized by $y^I=(x^i,t)$.  Using these coordinates, 
consider 
\eqn\eeloopone{T(y)=\Gamma_I\cdot y^I}
(where $\Gamma_I$ are now the $16\times 16$ gamma matrices thought of as 
maps between the two inequivalent irreducible spinor representations of 
the $SO(10)$ rotation group, $\Gamma_I:\CS_+ \rightarrow \CS_-$).  This loop 
in the space of configurations indeed satisfies our requirements.  It is 
topologically stable, because now the tachyon wraps $S^9$ at infinity once 
around the non-contractible $S^9$ in the vacuum manifold $U(16)$ (recall again 
that $\pi_9(U(16))=\Z$).  
Thus, despite our ignorance about the overall normalization factor, the  
family \eeloopone\ will indeed flow to a certain topologically non-trivial 
family of configurations.  This family is asymptotic to the supersymmetric 
vacuum at $t \rightarrow\pm\infty$, and by construction passes through the 
D0-brane configuration at $t=0$.  

In fact, the proper framework for understanding the non-contractible loop 
\eeloopone\ in the configuration space is K-theory.  Even though at each 
$t$ the Chan-Paton bundles of D9-branes and D$\bar 9$-branes are isomorphic 
(and given by the $SO(9)$ spinor bundle $\CS$), 
they wrap the extra dimension $t$ in a 
topologically nontrivial way, and span the non-isomorphic $SO(10)$ spinor 
bundles $\CS_+$ and $\CS_-$ (in accord with the fact that the $16\times 16$ 
gamma 
matrices of $SO(10)$ in \eeloopone\ provide a map $\CS_+\rightarrow \CS_-$).  
Thus, the Chan-Paton bundles of the whole one-parameter family of D9-D$\bar 9$ 
pairs represent a non-trivial element in the K-theory group of the extended 
manifold parametrized by $(x^i,t)$.  As an element of K-theory, the 
topological charge that stabilizes \eeloopone\ can be physically identified 
as one of the RR charges of Type IIB theory (namely, the D-instanton charge).  

Hence, we conclude that 

\item{(1)} the configuration space of Type IIB string theory 
has a non-contractible loop (supported by a topological charge that takes 
values in K-theory), and 

\item{(2)} the D0-brane of Type IIB string represents the D-sphaleron at the 
top of the potential barrier traversed by the loop.   

It should be pointed out that two important assumptions enter into this
conclusion. First, we have not defined from first principles, such as string 
field theory, what we understand by the configuration space of Type II string 
theory.  Instead, we are using the explicit construction of the D-sphalerons, 
in conjunction with the existence of RR charges as implied by K-theory, to 
{\it deduce\/} that the appropriately defined configuration space supports 
a non-contractible loop.  This configuration space contains all perturbative 
string configurations, plus the configurations of all possible sets of 
D-brane configurations (and possibly more). A priori, we cannot rule out the 
possibility that there is some yet to be understood part of the
configuration space which makes the above loop contractible.  However, 
this possibility seems unlikely, since the existence of a non-contractible 
loop in the configuration space follows from a topological argument:  
the loop is non-contractible because it carries a non-trivial K-theory class 
(essentially, one unit of the D-instanton charge).  As long as RR charges 
are conserved in the theory, it will not be possible to shrink the loop to 
a point.  

Second, the presence of a non-contractible loop only implies the existence of 
a sphaleron solution if the configuration space is compact. Pure Yang-Mills 
theory has non-contractible loops, but the non-compactness of configuration 
space generated by scale transformations forbids the existence of finite size 
sphaleron solutions.  We are assuming that in string theory, the string scale 
cuts off this source of noncompactness, and that the resulting object is the 
same as found by quantizing open strings with Dirichlet boundary conditions.   

Our conclusions can be easily generalized to the configuration space of 
extended configurations that fall off at infinity in directions normal to 
an extended hypersurface in space.  Just as in the case of the Type IIB 
D0-brane, one can interpret all the unstable Type IIB D$2p$-branes with $p>0$ 
as D-sphalerons, and deduce the the existence of a non-contractible loop in 
the corresponding configuration spaces of extended configurations.  

\newsec{D-Sphalerons in Type IIA Theory}

We now turn to a discussion of the interpretation of the Type IIA D-instanton.

Just as a non-contractible loop in the space of finite energy IIB
configurations implied the existence of the D0-sphaleron, a non-contractible
loop in the space of finite action IIA Euclidean histories gives rise
to a D-instanton with a single negative mode.  To exhibit the non-contractible 
loop, we proceed in parallel to the IIB discussion, now starting with 32 
unstable D9-branes.  Introducing the parameter $t$ and SO(10) gamma matrices 
$\Gamma_i$, the non-contractible loop is given by
\eqn\Tinst{T= \sum_{i=1}^{10} \Gamma_i x^i + \Gamma_{11} t.}
The loop gives a nontrivial element of $\pi_{10}(U(32)/ U(16) \times
U(16))$.  We identify the halfway point of the loop at $t=0$ with the 
IIA D-instanton.

Thus, the reason for the existence of  the IIA D-instanton is not
instability of the vacuum, rather it is required by the nontrivial
topology of the space of histories  in IIA string theory.  The topological 
charge that makes \Tinst\ stable corresponds to a non-trivial element of 
K-theory, with a very interesting physical interpretation:  in K-theory, 
this topological charge can be identified as the RR D$(-2)$-brane charge.  
Recall that in Type IIA string theory, there is a RR ten-form $F_{10}$ 
(related to the cosmological constant in massive Type IIA theory), which 
couples to the D$8$-brane; formally, the magnetic dual of the D$8$-brane 
should be a D$(-2)$-brane, a concept that is indeed very hard to understand 
in physical terms.  Here we have found a natural physical role of the 
D$(-2)$-brane 
charge (if not the D$(-2)$-brane), as the topological charge responsible for 
the non-contractible loop in the space of Type IIA histories.  In a formal 
sense, one can think of the D$(-2)$-brane as an ``object'' localized in the 
extra dimension of the one-parameter family of histories traversing this 
non-contractible loop.  

So far, our discussion of Type IIA theory has been focused on interpreting the 
Type IIA D-instanton, and therefore we were looking at the space of 
Euclidean histories.  Similar arguments can also be used to analyze the 
configuration space of Type IIA theory.  Interpreting {\it nine\/} of the 
eleven dimensions in \Tinst\ as space dimensions, and the remaining two as 
extra parameters, we can view \Tinst\ as a two-parameter family of string 
configurations that corresponds to a non-contractible two-sphere in the 
configuration space of Type IIA string theory.  The corresponding sphaleron 
at the far pole of this non-contractible $S^2$ is easy to find by setting 
the two parameters representing the $S^2$ in \Tinst\ equal to zero.  The 
sphaleron configuration that we obtain,
\eqn\eedoantido{T=\pmatrix{\vec\Gamma\cdot\vec x&0\cr & \cr
0&-\vec\Gamma\cdot\vec x\cr}}
(with $\vec x$ describing the nine space dimensions and $\vec\Gamma$ the Gamma 
matrices of $SO(9)$), was already encountered in a different context in 
\eedectwo , and describes a coincident D0-D$\bar 0$ pair.  

The identification of the sphaleron in Type IIA configuration space as a 
D0-D$\bar 0$ pair nicely agrees with the expected counting of negative modes.  
The D0-D$\bar 0$ system has a complex tachyon, from the open string stretching 
between the D0 and the D$\bar 0$-brane.  This tachyon gives two real negative 
modes, precisely as expected from the sphaleron on the far pole of an $S^2$.  

\newsec{Topology of Configuration Space in String Theory}

We have seen how to relate a single D-sphaleron to a non-contractible loop
in configuration space.  This loop is non-contractible because the 
corresponding one-parameter family of string configurations carries a 
topological charge in K-theory, even though each individual configuration 
carries zero charge.  This structure 
clearly generalizes to multi-parameter families of string configurations.  
Looking back at \eehomvone\ and \eehomvtwo\  (or, more abstractly, invoking 
Bott periodicity in K-theory), we can generalize the construction of the 
section~3, and demonstrate that the string configuration space of 
Type IIB (IIA) string theory contains non-contractible spheres $S^k$ of 
arbitrarily large odd (even) dimension $k$.  In turn, each non-contractible 
$S^k$ implies the existence of a sphaleron solution (with exactly $k$ 
negative modes), at the pole of $S^k$ opposite to the vacuum.  What is the 
physical interpretation of such higher sphaleron solutions?  

In this section we show that these higher sphalerons do not represent novel 
solutions; rather, they can be interpreted as multiple coincident 
D0-sphalerons of the previous section.  We will demonstrate explicitly that we 
recover the correct counting of negative modes on $k$ D-sphalerons.  

\subsec{Higher non-contractible spheres in the IIB configuration space}

\fig{Non-contractible $2n-1$-sphere in the configuration space and the 
corresponding  $n$ sphaleron configuration.}{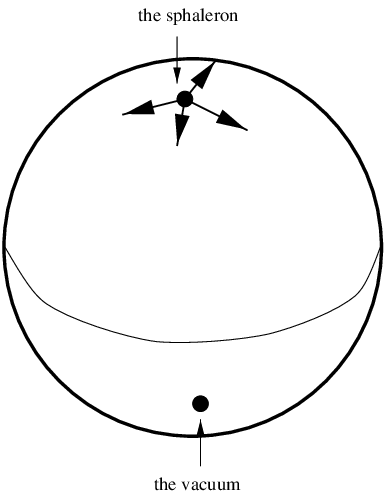}{2truein}

We begin with a  concrete example relating two coincident D0-branes in IIB
to a non-contractible $S^3$ in the space of finite energy nine dimensional
field configurations.   We have seen that a single
D0-brane can be represented as the point $t=0$ on the loop 
$T= \Gamma_i x^i + \Gamma_{10} t$,
where $i=1 \ldots 9$, and $\Gamma_i$ are $16 \times 16$ 
$SO(10)$ gamma matrices.   To 
represent two D0-branes, we introduce three parameters $t_1, t_2, t_3$, and
define a non-contractible $S^3$ in terms of $SO(12)$ gamma matrices by
\eqn\twobranes{T= \tilde{\Gamma}_i x^i + \tilde{\Gamma}_{10}t_1 + 
\tilde{\Gamma}_{11} t_2 +  \tilde{\Gamma}_{12} t_3.}
  Choosing a convenient
representation for $\tilde{\Gamma}_i$, this becomes
\eqn\Ttwo{T=\pmatrix{\Gamma_i x^i + \Gamma_{10}t_1
&(t_2 -  t_3)\, 1_{16} 
 \cr
  & \cr (t_2 +  t_3)\, 1_{16}
&-(\Gamma_i x^i + \Gamma_{10}t_1)\cr}.}
It is evident that the ``far pole''  of the $S^3$ at $t_1=t_2=t_3=0$, 
, as depicted in fig.~4, represents two coincident D0-branes.  

On the two D0-branes we expect to find $2^2=4$ negative modes.  Three 
negative modes arise from motion on the $S^3$, {\it i.e.} 
$\delta T = \tilde{\Gamma}_{9+i} \delta t_i$ $(i=1,2,3)$.  The final negative
mode arises from motion on the non-contractible $S^1$ as for a single
D0-brane:
\eqn\deltaTtwo{T+\delta T =\pmatrix{\Gamma_i x^i + \Gamma_{10} \delta t
& 0 
\cr  &  \cr
0 
&-\Gamma_i x^i + \Gamma_{10} \delta t\cr}.}
So we indeed correctly reproduce the 4 negative modes known to exist from the 
quantization of open strings.  

This procedure can be directly generalized to construct a non-contractible 
$S^n$ for all odd values of $n$, whose existence is suggested by Bott 
periodicity of the homotopy groups \eehomvone .  The generalization involves 
an interesting subtlety, which is best illuminated as follows.  To simplify 
the argument, consider the unstable D0-brane as a real kink on the worldsheet 
of a coincident D1-D$\bar 1$ pair along its space-like dimension $x$.  The 
non-contractible $S^1$ in the configuration space is described by the stable 
vortex on the two-manifold spanned by $(x,t)$, where $t$ is the parameter 
along the loop.  At each fixed $t$, we have one D1-D$\bar 1$ pair.  Similarly, 
the non-contractible $S^3$ discussed above corresponds to a point-like defect 
on a four-manifold spanned by $(x,t^1,t^2,t^3)$; to construct such a defect, 
we need a family consisting of {\it two\/} D1-D$\bar 1$ pairs at each $t^i$.  
This procedure can be iterated; in each step, as we add two more 
parameters $t^{2k},t^{2k+1}$, the $\Gamma\cdot y$ representation of the 
non-contractible $S^{2k+1}$ requires doubling the number of D1-D$\bar 1$ 
pairs.  Thus, the non-contractible $S^{2k+1}$ requires a family of 
$2^k$ D1-D$\bar 1$ pairs parametrized by $t^1,\ldots t^{2k+1}$.  Notice 
that the number of D1-D$\bar 1$ pairs grows exponentially with growing $k$.  

This construction certainly leads to a non-contractible $S^{2k+1}$ in the 
configuration space, and one might be tempted to identify the configuration 
at $t^i=0,i=1,\ldots 2k+1,$ as the corresponding sphaleron.  However, a small 
puzzle immediately appears.  While it is easy to show that the configuration 
at $t^i=0$ is given by 
\eqn\cfgzero{T=x\cdot 1_{2^k}}
and consists therefore of $2^k$ coincident D0-sphalerons, it is also 
straightforward to see that for $k>1$ such a configuration has too many 
negative modes to represent the sphaleron at the far pole of $S^{2k+1}$, 
whose number of negative modes should grow linearly and not exponentially 
with $k$.  

This puzzle is resolved by the following observation.  One can certainly use 
the $\Gamma\cdot y$ construction to conveniently construct the non-trivial 
element of $\pi_{2k+1}(U(N))$, but the number $N=2^k$ of D1-D$\bar 1$ 
pairs needed in this construction is not the smallest one possible; in fact, 
it is deeply inside the stability regime.  In order to identify the 
sphaleron, we have to minimize the energy of the configuration at the far pole 
of the $S^{2k+1}$, and for that we need to use the smallest possible number of 
D1-D$\bar 1$ pairs allowed by the stability bound.  This bound requires 
$N\geq k+1$ pairs to properly accommodate $\pi_{2k+1}$!  On this minimal 
number $k+1$ of D1-D$\bar 1$ pairs, the sphaleron at $t^i=0$ corresponds to 
$k+1$ coincident D0-branes.  

Thus, we claim that the sphaleron on the far pole of the non-contractible 
$S^{2n-1}$ is given by $n$ coincident unstable D0-branes.  It is now easy to 
see that the count of the number of negative modes indeed works as expected.  
The configuration of $n$ coincident D0-branes exhibits $n^2$ negative modes, 
corresponding to the open-string tachyon in the adjoint of $U(n)$.  Just like 
in the case of $n=2$ discussed explicitly above, it is important to realize 
that the system of $n$ D0-branes contains subsystems of $p<n$ D0-branes that 
sit at the far pole of $S^{2p-1}$ for all $p=1,\ldots n-1$.  Motion on each 
$S^p$ is associated with $p$ negative modes.  Thus, the total number of 
negative modes is 
\eqn\negnum{ 1+3+5+ \cdots + 2n-1 = n^2,}
as expected.

An analogous counting of negative modes goes through for configurations
of coincident D$2p$-branes, including configurations which include 
branes of different dimensionalities.  It is a satisfying consistency check
that in all these cases, we reproduce the same spectrum of negative modes as 
arises from the quantization of open strings on non-BPS D-branes.

We are therefore led to conclude that 
\item{(1)} the configuration space of Type IIB string theory has a 
homotopy structure which is at least as complicated as that of the infinite 
unitary group $U(N)$, $N\rightarrow\infty$: $\pi_k$ of the configuration 
space is non-trivial for all odd $k$; 

\item{(2)} similarly, the configuration space of Type IIA string theory 
has a homotopy structure at least as complicated as that of an infinite 
Grassmannian, $U(2N)/U(N)\times U(N)$, with all $\pi_{2k}$ nontrivial.   

\subsec{Connection to K-theory}

Our discussion so far has involved specific examples of D-brane sphalerons and 
non-trivial homotopy groups of the configuration space of Type II string 
theory in flat $\R^{10}$.  It is perhaps worth stressing that the connection 
between D-sphalerons, K-theory, and the non-trivial homotopy groups of the 
string configuration space is quite universal, and our results naturally 
generalize to more complicated cases, including compactifications and 
orientifolds.  

Consider any compactification of Type II or Type I theory.  For simplicity, 
we will discuss the case of Type IIB theory compactification on $X$, but the 
generalization to other theories is straightforward.  Stable D-branes on $X$ 
are classified by elements of the (reduced) K-theory group $K(X)$, which in 
turn can be identified as the group of equivalence classes of pairs of 
Chan-Paton bundles $(E,F)$ on a number of spacetime-filling D9-D$\bar 9$ 
pairs wrapping $X$.  The equivalence relation corresponds to creation and 
annihilation of pairs from/to the vacuum.  

Imagine now an $n$-parameter family of D9-D$\bar 9$ pairs, with Chan-Paton 
bundles $(E(t),F(t))$.  In our discussion so far, the parameters 
$t=(t^1,\ldots t^n)$ were coordinates on an $S^n$, but one can consider a 
general $n$-manifold $Y$ of parameters.  For any fixed 
$t$, $(E(t),F(t))$ defines an element $\alpha(t)$ of $K(X)$, and the whole 
family defines an element of $K(X\times Y)$.  Even if $\alpha(t)$ is 
trivial for each $t$, the element of $K(X\times Y)$ defined by the whole 
family can be non-trivial.  When this is so, the family represents a 
non-contractible manifold $Y$ in the configuration space of the theory in 
the vacuum sector.  

Thus, there is an intimate relation between the homotopy structure of the 
string configuration space on $X$ and K-theory groups of $K(X\times Y)$ 
for various $Y$.  Since the latter are related to the spectrum of D-brane 
charges on $X$, the homotopy structure of the configuration space is closely 
related to the stable D-brane spectrum on $X$.  Assuming that the 
configuration space is sufficiently compact, the non-trivial elements of the 
homotopy groups in turn imply the existence of corresponding D-sphalerons.  

\newsec{Tachyon Condensation and Massive Type IIA Vacua}

As mentioned in section 2, we have been imposing certain restrictions
in our study of tachyon configurations on unstable D9-branes in IIA.   We 
chose to start with an even number $2N$ of D9-branes, and assumed that 
tachyon condensation Higgsed the gauge group according to $U(2N) \rightarrow 
U(N) \times U(N)$.  This symmetry breaking pattern with an even number of 
unstable D9-branes arose in \phk , where it was found to be directly related 
to K-theory and the classification of all D-brane charges in Type IIA theory.  
However, the role of other Higgs patterns, and configurations with an odd 
number of unstable D9-branes was left somewhat mysterious in the analysis of 
\phk .  

In this section our conditions will be relaxed:  we allow an arbitrary number 
$N$ of D9-branes, as well as the general Higgsing pattern $U(N) \rightarrow 
U(k) \times U(N-k)$.  We will be led to an interpretation of these 
configurations in terms of vacua with non-vanishing flux for the RR 10-form 
$F_{10}$. 

Let us recall some aspects of vacua with $F_{10}$ 
flux \refs{\polchinski, \polchstrom}.  Including the
non-propagating field
$F_{10}$ in type IIA supergravity leads to  so-called massive IIA
supergravity \romans.  The field equations of this theory admit solutions
with constant $F_{10}$, and which preserve all $32$ supersymmetries. 
In string theory it has been argued \refs{\polchstrom,\bergshoeff,\green} that such vacua exist only for
a discrete set of fluxes, $\nu\equiv ^*\!\!\!F_{10} = n \mu_8$, where $\mu_8$ is the 
tension of a BPS D8-brane.  D8-branes play the role of domain walls
between distinct vacua, with $\nu$ jumping by $\mu_8$ upon crossing
a D8-brane.  We also remark that the massive IIA theory has a cosmological 
constant via $S \sim \int d^{10}x \sqrt{-g}\, \nu^2$, 
and that the theory cannot be obtained from the dimensional reduction
of any known eleven-dimensional theory.

To connect the above facts to our discussion, we first examine the simple
case of a single unstable D9-brane.  On the worldvolume of the D9-brane there
is a neutral tachyon $T$, whose potential $V(T)$ is assumed to be of the
standard double-well form, with a local maximum at $T=0$ and minima
at $T= \pm T_0$.  As in \phk, it is conjectured that a BPS D8-brane is
represented by a kink configuration; {\it i.e.} $T=f(x_9) x_9$ describes
a D8-brane at $x_9=0$, where  $f(x_9)$ is a smooth function behaving as 
$T_0/|x_9|$ for large $|x_9|$.  The kink will carries the RR charge of a
D8-brane given  that on the D9-brane there exists a coupling to
 the RR 9-form potential $C_9$ of the form \refs{\phk,\sena,\bcr}
\eqn\WZ{S = { \mu_8 \over 2 T_0} \int  dT \wedge C_9.}
There is no straightforward way to directly compute the coefficient of
this term, since the presence of $T_0$ in the denominator shows that 
it depends on unknown details of the tachyon potential.
We have chosen   the coefficient
so that the kink carries the charge of a single D8-brane as in \bcr.

Now consider the homogeneous tachyon configurations $T=0$ and $T=\pm T_0$,
and imagine an adiabatic process in which the tachyon is taken from one
such solution to another.   The quadratic term for $F_{10}$ along with
the coupling \WZ\ yield the field equation
\eqn\Feqtn{d ^*\!F_{10}= { \mu_8 \over 2 T_0} dT.} 
Hence in taking the tachyon from one minimum, $T=-T_0$, to the other, 
$T=+T_0$, we find that $F_{10}$ changes by $\Delta \nu =  \mu_8$.
Given the previous quantization condition for $\nu$ in the massive IIA
vacua, it is natural to conclude that in the process of shifting the 
tachyon we have moved from one massive IIA vacuum to an adjacent one.  
In this interpretation, a D8-brane, described as a kink,
 indeed represents a domain wall between distinct massive IIA vacua.  
On the other hand, if we adiabatically take the tachyon from $T=-T_0$ to
the unstable local maximum at
$T=0$, we find  $\Delta \nu =  \mu_8/2$ which, perhaps
surprisingly, forces us to admit values of $\nu$ not included among
the massive IIA vacua.  That is, we learn that in order to respect the 
quantization of $\nu$ after tachyon condensation, we must
have that a single D9-brane with vanishing $T$ can only exist in the
 presence of half odd integer units of flux: $\nu = (n+1/2)\mu_8$. 

The foregoing analysis is easily generalized to the case of $N$ unstable
D9-branes.  We assume that $V(T)$ has minima of the form 
\eqn\gentach{T= T_0 \pmatrix{1_k  & 0 \cr & \cr 0 & - 1_{N-k} \cr},}
and that on the D9-branes there exists a coupling
\eqn\genWZ{S = { \mu_8 \over 2 T_0} \int {\rm Tr}(dT) \wedge C_9.}
Adiabatic variation of $T$ then gives 
$\Delta \nu = {1 \over 2} {\mu_8 \over T_0} \Delta {\rm Tr}(T)$.  
By moving between different minima, one can reach values for 
$\nu$ corresponding to any given massive IIA vacuum.  For  $N$ even,
it is consistent to take $\nu=0$ at $T=0$, and also after tachyon
condensation to the traceless configuration $k=N/2$.  This is what has
been assumed in the bulk of this paper.  But for  $N$ odd, consistency
with the quantization condition requires one to include half odd integer
units of flux at $T=0$.  

One might be suspicious of the need to introduce half odd integer units
of flux, given what was said about the difficulty in computing the
coefficient of the term \WZ.  Perhaps the assumed coefficient is incorrect
by a factor of two, so that a kink truly represents two D8-branes.  To
allay such suspicions, we will compute the spectrum of fermion zero modes
on the kink, and see that we obtain a single $8+1$ dimensional Majorana 
fermion, modulo one assumption,
as we should if the kink represents a single D8-brane. 

The computation is closely related to one performed in \senspinors, which 
yielded the fermion zero modes on a Type I D0-brane regarded as a kink on a 
D$1$-D$\bar 1$ pair.  On an unstable D9-brane are two Majorana-Weyl fermions
of opposite chiralities, $\psi_{\pm}$.  We take these to  couple to the 
tachyon at quadratic order through an action of the form
\eqn\feract{S = \int \! d^{10}x\,\left\{  {i \over 2} f_1(T)
[\psi_+^T \Gamma^0 \Gamma^\mu \partial_\mu \psi_+
+ \psi_-^T \Gamma^0 \Gamma^\mu \partial_\mu \psi_-]
+ f_2(T) \psi_+^T \Gamma^0 \psi_- \right\}.}
$\Gamma^\mu$ are purely imaginary $SO(9,1)$ gamma matrices. $f_{1,2}(T)$
are functions of $T$ and its derivatives.  The action is restricted by a 
$\Z_2$ symmetry \sena\ which flips the sign of $T$ along with one of the 
fermions;  the symmetry requires $f_1$ to be an even function of $T$, and 
$f_2$ to be an odd function of $T$. The couplings are also restricted by a 
non-linearly realized supersymmetry acting on the fermion fields as discussed 
in \refs{\senpuz,\yoneya}. It is not clear whether this fact is compatible 
with the last term in \feract, or with the analogous term in \senspinors, 
although the equations of motion which follow from \feract\ appear to be 
compatible with those in \yoneya\ to lowest order. This question deserves 
closer scrutiny, for now we will assume that \feract\ is correct and proceed. 

For the tachyon background we take a kink located at $x_9=0$.  As with the
tachyon potential $V(T)$, there is no systematic way to calculate the
functions $f_{1,2}$.    Our
main assumption is that for  a kink background $f_2/f_1$ goes to a 
nonzero  constant  --- which can be taken to be positive --- 
 for large $x_9$, and hence to a negative constant for
large $-x_9$ as the tachyon moves from one minimum of its potential
to the other. 
 Fermion zero modes are obtained from
normalizable solutions to the Dirac equation which depend only on $x_9$. 
Defining the linear combinations
\eqn\chidef{\chi_\pm = \psi_+ \pm \psi_-,}
the Dirac equation is found to be 
\eqn\dirac{\partial_9 \chi_\pm = - \left[ {1\over 2}
{\partial_9 f_1 \over f_1} \pm i {f_2 \over f_1} \Gamma^9\right] \chi_\pm.}
The solutions are
\eqn\dirsol{\chi_\pm = f_1^{-1/2}
\exp{\left[ \mp i \int_0^{x_9} \!\! dx_9' \, {f_2 \over f_1} \Gamma^9 \right]} 
\chi_\pm^{(0)},}
where $\chi_\pm^{(0)}$ are constant spinors.  Given the assumed behavior
of $f_2/f_1$, normalizability requires
\eqn\norm{\Gamma^9 \chi_+^{(0)}=-i\chi_+^{(0)}, \quad\quad
\Gamma^9 \chi_-^{(0)}=+i\chi_-^{(0)}.}

With these projections, the spectrum of fermion zero modes is that of a
single $8+1$ dimensional Majorana fermion. Thus we have verified that a
kink on an unstable D9-brane represents a single BPS D8-brane, which in 
turn requires that an odd number of unstable D9-branes be accompanied by
 half odd integral units of 10-form flux.  

In closing this section, we point out that according to 
\refs{\alvarez,\wittenglobal} an $8+1$ dimensional theory with an odd number
of Majorana fermions potentially suffers from a global gravitational 
anomaly.  In the present case, the $8+1$ dimensional theory on the kink
was obtained by starting from an anomaly free $9+1$ dimensional theory, which
indicates that the anomaly should cancel through some global version of
anomaly inflow. This anomaly problem has recently been addressed in
\moorwit.   

\newsec{Conclusions and Outlook}

In this paper, we have established the existence of finite-energy sphalerons 
in perturbative string theory, and identified them with the previously 
studied unstable D-branes.  Thus, the unstable D-branes are legitimate objects 
in string theory, tied to the existence of a complicated homotopy structure of 
the configuration space of the theory and the existence of RR charges (or, 
more generally, charges in K-theory) in the ``right'' dimensions.  As 
mentioned earlier,  it is clear from the connection to K-theory that the 
structure uncovered in this paper is very universal, and a much richer 
spectrum of D-sphalerons is to be expected upon compactification.  It will be 
interesting to unravel the implications of such D-sphalerons in more 
complicated situations. 
 
Our construction of D-sphalerons was perturbative in $g_s$.  Unlike their 
RR-charged BPS counterparts, D-sphalerons do not carry any conserved quantum 
numbers, and there is no a priori reason to expect that they survive as 
pronounced objects beyond the regime of weak string coupling.  Therefore, our 
conclusions about the structure of the configuration space are strictly valid 
at small $g_s$ only. Nonetheless, since the existence of D-sphalerons is 
protected by the existence of BPS RR-charges (and is therefore topological 
in nature, related to K-theory), it seems natural to expect that at least some 
aspects of the sphalerons will survive even at large $g_s$.  In principle, 
one can ask whether the homotopy structure of the string configuration space 
can be recovered in a dual description of a given theory.  It is amusing that 
infinite Grassmannians appeared previously in the string theory literature  
in early attempts to go beyond perturbation theory, where they played the role 
of the universal moduli space of all Riemann surfaces (including surfaces of 
infinite genus) \infgenus .  

Our construction sheds light on the existence of the elusive D$(-2)$-brane 
of Type IIA string theory, which couples to $F_{10}$ and therefore is 
important for issues that have to do with the cosmological constant.  The 
D$(-2)$-brane charge was found responsible for the existence of a 
non-contractible loop in the space of Type IIA histories in $\R^{10}$.  

Although the Type IIA D-instanton -- being an example of a D-sphaleron -- does 
not cause false vacuum decay, of the supersymmetric vacuum of IIA theory,
the closely related Euclidean Schwarzschild instanton will lead to false
vacuum decay  of $M$ theory on $\R^{10}\times S^1$ with the anti-periodic
choice of spin structure on the $S^1$ following the analysis of \ewkk. 
This process has interesting generalizations to other non-supersymmetric
string compactifications \hhktwo. 

Finally, we have not yet explored the physical implications of D-sphalerons
in string theory. 
In field theory, sphalerons represent solutions at the top of a finite-energy 
barrier that can be classically overcome under favorable circumstances. In
certain regimes they provide the leading semi-classical contribution to
certain processes such as baryon number violation in the standard model. 

At finite temperature, one can create field-theory sphalerons 
because they are soft and large objects, relatively easy to create by a large 
number of soft quanta in the thermal ensemble.  
In high-energy scattering processes, on the other hand, it might be 
difficult to create a soft large sphaleron by scattering a few very energetic 
quanta, and it has been argued in field theory that baryon-mediated processes 
are not enhanced \dlsfp .  

In string theory, D-sphalerons are objects that have a hard core under a 
stringy halo.  Therefore, one can expect that -- unlike in field theory -- 
the stringy D-sphalerons could play an important role in high-energy 
scattering processes.  On the other hand, their possible role at finite 
temperatures seems more obscure.  At small values of the string coupling, 
the mass of the sphalerons is proportional to $\sqrt{\alpha'}/g_s$, and before 
we reach that energy regime in the thermal ensemble, we encounter the Hagedorn 
transition.

\bigskip\medskip\noindent
We would like to thank Eric Gimon, Ruth Gregory, Chris Hull, Emil Martinec, 
Djordje Minic, Albert Schwarz, Steve Shenker, and Edward Witten for helpful 
conversations. 
The work of J.H. is supported in part by NSF Grant No.\ PHY 9901194.  
The work of P.H. is supported in part by a Sherman Fairchild Prize 
Fellowship, and by DOE Grant No.\ DE-FG03-92-ER~40701.  The work of P.K. is 
supported in part by NSF Grant No.\ PHY 9901194 and by  NSF Grant No.\ 
PHY94-07194. 

\listrefs
\end